\documentstyle[prl,aps,epsf,floats]{revtex}
\begin{document}
\draft

\twocolumn[\hsize\textwidth\columnwidth\hsize\csname @twocolumnfalse\endcsname

\title{High-frequency spin valve effect in ferromagnet-semiconductor-ferromagnet 
structure based on precession of injected spins}
\author{A.M. Bratkovsky and V.V. Osipov }
\address{
Hewlett-Packard Laboratories, 1501 Page Mill Road, 1L, Palo Alto, CA 94304}
\date{November 19, 2003}
\maketitle

\begin{abstract}

New mechanism of magnetoresistance, based on tunneling-emission 
of spin polarized electrons from ferromagnets (FM) into semiconductors (S) and 
precession of electron spin in the semiconductor layer under external 
magnetic field, is described. The FM-S-FM structure is considered, which includes 
very thin heavily doped ($\delta$-doped) layers at FM-S interfaces. At
certain parameters  
the structure is highly sensitive at room-temperature to variations of the field 
with frequencies up to 100 GHz. The current oscillates with the field, 
and its relative amplitude is determined  only by the spin
polarizations of FM-S junctions at relatively large bias
voltage. 

\end{abstract}
\pacs{72.25.Hg, 72.25.Mk}

\vskip2pc]

\narrowtext

Manipulation of an electron spin may lead to breakthroughs in solid state
ultrafast scalable devices \cite{Wolf}. Spintronic effects, like giant and
tunnel magnetoresistance (TMR) are already widely used in read-out devices
and non-volatile memory cells \cite{Wolf,GMR}. Theory of TMR junctions has
been considered in Refs. \cite{Slon,Brat}. A large ballistic
magnetoresistance of Ni and Co nanocontacts was reported in Refs.~\cite{Garc}%
. The injection of spin-polarized carriers into semiconductors provides a
potentially powerful mechanism for field sensing and other applications,
which is due to relatively large spin-coherence lifetime of electrons in
semiconductors \cite{Wolf,Awsh}. Different spintronic devices, including
magnetic sensors, are considered in detail in \cite{Wolf}. The efficient
spin injection into nonmagnetic semiconductors has been recently
demonstrated from ferromagnets \cite{Ferro,Jonk} and magnetic semiconductors 
\cite{MSemi}. Conditions for efficient spin injection have been discussed in
Refs. \cite{Rash,BO03}. Spin diffusion and drift in electric field have been
studied in Refs.~\cite{aronov76pik,flat}.

In this paper we study a new mechanism of magnetoresistance, operational up
to 100 GHz frequencies. We consider a heterostructure comprising a $n-$type
semiconductor ($n-$S)\ layer sandwiched between two ferromagnetic (FM)\
layers with ultrathin heavily $n^{+}-$doped ($\delta -$doped) semiconducting
layers at the FM-S interfaces. Magnetoresistance of the heterostructure is
determined by the following processes: (i) injection of spin polarized
electrons from the left ferromagnet through the $\delta -$doped layer into
the $n-$S layer; (ii) spin ballistic transport of spin polarized electrons
through that layer; (iii) precession of the electron spin in an external
magnetic field during a transit through the $n-$S layer; (iv) variation of
conductivity of the system due to the spin precession.

There are known obstacles for an efficient spin injection in FM-S
structures. A Schottky barrier with a height $\Delta \gtrsim 0.5$ eV usually
forms in a semiconductor near a metal-semiconductor interface \cite{Sze}.
The energy band diagram of a thin FM-S-FM structure looks as a rectangular
potential barrier of a height $\Delta $ and a thickness $w$. Hence, the
current through the FM-S-FM structure is negligible when $w\gtrsim 30$ nm.
To increase a spin injection current, a thin heavily $n^{+}-$semiconductor
layer between the ferromagnet and semiconductor should be used \cite
{Jonk,BO03}. This layer sharply decreases the thickness of the Schottky
barriers and increases their tunneling transparency \cite{Sze}. Recently
an efficient injection was demonstrated in FM-S junctions with a thin $%
n^{+}-$layer \cite{Jonk}.

We consider a heterostructure, Fig.~1, where left (L) and
right (R) $\delta-$doped layers satisfy the following optimal
conditions \cite{BO03}: 
the thickness $l^{L(R)}\lesssim 2$ nm, the donor concentration $%
N_{d}^{+}\gtrsim 10^{20}$cm$^{-3},$ $N_{d}^{+}(l^{L})^{2}\simeq 2\varepsilon
\varepsilon _{0}(\Delta -\Delta _{0}+rT)/q^{2},$ and $N_{d}^{+}(l^{R})^{2}%
\simeq 2\varepsilon \varepsilon _{0}(\Delta -\Delta _{0})/q^{2},$ where $%
\Delta _{0}=E_{c}-F$, $F$ is the Fermi level in the equilibrium (in the left
FM), $E_{c}$ the bottom of semiconductor conduction band, $r\simeq 2-3,$
and $T$ the temperature (we use the units of $k_{B}=1)$. The value of $%
\Delta _{0}$ and the relevant profile of $E_{c\text{ }}(x)$ can be set by
choosing $N_{d}^{+}$, $l^{L(R)},$ and a donor concentration, $N_{d},$ in the 
$n-$semiconductor. The energy diagram of such a FM$-n^{+}-n-n^{+}-$FM\
structure is shown in Fig.~1. Importantly, there is a shallow potential well
of depth $\approx rT$ next to the left $\delta -$spike. Presence of this
mini-well allows to\ retain the thickness of the left $\delta -$barrier
equal to $l^{L}\lesssim l_{0}$ and its tunneling transparency high for the
bias voltage up to $qV_{L}\simeq rT$. The $\delta -$spike is transparent for
tunneling when $l^{L(R)}\lesssim l_{0}=\sqrt{\hbar^2/[2m_{\ast }(\Delta -\Delta
_{0})]}$, where $m_{\ast }$ is the effective mass of electrons in the
semiconductor. However, when $w\gg l_{0},$ only electrons with
energies $E\geq E_{c}=F+\Delta _{0}$ can overcome the barrier $\Delta _{0}$
due to thermionic emission \cite{BO03}. We assume $w\gg \lambda $, $\lambda $
being the electron mean path in a semiconductor, so one can consider the
FM-S junctions independently. 
%%  FIGURE 1    %%%%%%%%%%%%%%%%%%%%%%%%%%%%%%%%%%%%%%%%%%%%%%%%%%%%%%%%%%%%
\begin{figure}[t]
\epsfxsize=3.in \epsffile{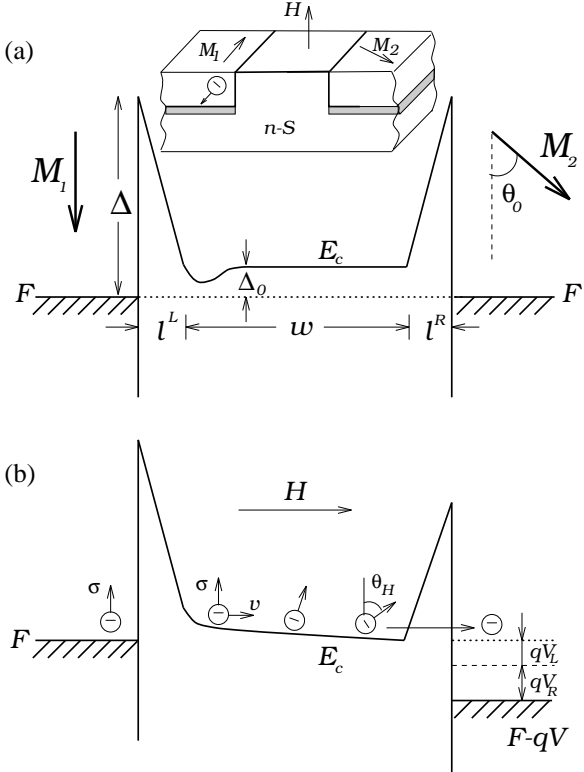}
\caption{Energy diagram of a the FM-S-FM heterostructure with $\protect%
\delta -$doped layers in equilibrium (a) and at a bias voltage $V$ (b), with 
$V_{L}$ ($V_{R}$) the fraction of the total drop across the left (right) $%
\protect\delta $-layer. $F$ marks the Fermi level, $\Delta $ the height, $%
l^{L(R)}$ the thickness of the left (right) $\protect\delta -$doped layer, $%
\Delta _{0}$ the height of the barrier in the $n-$type semiconductor (n-S), $%
E_{c}$ the bottom of conduction band in the n-S, $w$ the width of the n-S
part. The magnetic moments on the FM\ electrodes $\vec{M}_{1}$ and $\vec{M}%
_{2}$ are at some angle $\protect\theta _{0}$ with respect to each other.
The spins, injected from the left, drift in the semiconductor layer and
rotate by the angle $\protect\theta _{H}$ in the external magnetic field $H$%
. Inset: schematic of the device, with an oxide layer separating the
ferromagnetic films from the bottom semiconductor layer. }
\label{fig:fig1}
\end{figure}
We assume the elastic coherent tunneling, so that the energy $E$ and the
wave vector $\vec{k}_{\parallel }$ in the plane of the interface are
conserved, so the current density of electrons with spin $\sigma $ through
the left and right junctions, including the $\delta -$doped layers, can be
written as \cite{Duke,Brat,BO03} 
\begin{equation}
J_{\sigma }^{L(R)}=\frac{q}{h}\int dE[f(E-F_{\sigma
}^{L(R)})-f(E-F_{L(R)})]\int \frac{d^{2}k_{\parallel }}{(2\pi )^{2}}%
T_{k\sigma }^{L(R)},  \label{JRL}
\end{equation}
where $T_{k\sigma }$ is the transmission probability, $\ f(E)=[\exp
(E-F)/T+1]^{-1}$ the Fermi function, $F_{L}=F$ and $F_{R}=F-qV$ is the left
and right Fermi level, Fig. 1, $q$ the elementary charge, and integration
includes a summation with respect to a band index. We take into account the
spin accumulation in the semiconductor described by the Fermi functions with
the nonequilibrium quasi Fermi levels $F_{\sigma }$. The condition $%
\Delta _{0}=E_{c}-F>0$ means that the semiconductor is nondegenerate, so
that the total electron density, $n=N_{d},$ and a density of electrons with
spin $\sigma ,$ $n_{\sigma }$, are given by 
\begin{equation}
n=N_{c}\exp \left( -\frac{\Delta _{0}}{T}\right) =N_{d},\text{ }n_{\sigma }=%
\frac{N_{c}}{2}\exp \left( \frac{F_{\sigma }-E_{c}}{T}\right) ,  \label{nc}
\end{equation}
where $N_{c}=2M_{c}(2\pi m_{\ast }T)^{3/2}h^{-3}$ is the effective density
of states in the semiconductor conduction band and $M_{c}$ the number of the
band minima \cite{Sze}. The left (right) junctions are at $x=0$ ($w)$, so
that in Eq.~(\ref{JRL}) $F_{\sigma }^{L}=F_{\sigma }(0)$ and $%
F_{\sigma }^{R}=F_{\sigma }(w)$. The analytical expressions for $%
T_{k\sigma }^{L(R)}$ can be obtained in an effective mass approximation, $%
\hbar k_{\sigma }=m_{\sigma }v_{\sigma }$, where $v_{\sigma }$ is a velocity
of electrons with spin $\sigma $. The present interface barriers are opaque
at energies $E<E_{c}$. For energies\ $E\gtrsim E_{c}$ we can approximate the 
$\delta -$doped barrier by a triangular shape and find \cite{BO03} 
\begin{equation}
T_{k\sigma }^{L(R)}=\frac{16\alpha ^{L(R)}v_{\sigma x}^{L(R)}v_{x}^{L(R)}}{%
(v_{\sigma x}^{L(R)})^{2}+(v_{tx}^{L(R)})^{2}}\exp \left( -\eta \kappa
^{L(R)}l^{L(R)}\right) ,  \label{tran}
\end{equation}
where $E_{\parallel }=\hbar ^{2}k_{\parallel }^{2}/2m_{\ast }$, $%
v_{tx}^{L(R)}=\hbar \kappa ^{L(R)}/m_{\ast }$ the ``tunneling'' velocity, $%
v_{x}^{L(R)}=\sqrt{2(E-E_{c}-E_{\parallel })/m_{\ast }}$ and $v_{\sigma
x}^{L(R)}$ are the $x-$components of electron velocities in a direction of
current in the semiconductor and ferromagnets, respectively, $\kappa
^{L(R)}=(2m_{\ast }/\hbar ^{2})^{1/2}(\Delta +F-E+E_{\parallel
})^{3/2}(\Delta -\Delta _{0}\pm qV_{L(R)})^{-1}$, $\alpha
^{L(R)}=3^{-1/3}\pi \Gamma ^{-2}\left( \frac{2}{3}\right) \left[ \kappa
^{L(R)}l^{L(R)}\right] ^{1/3}\simeq 1.2\left[ \kappa ^{L(R)}l^{L(R)}\right]
^{1/3},$ $\eta =4/3$ (for a rectangular barrier $\alpha =1$ and $\eta =2$),
where $V_{L(R)}$ is the voltage drop across the left (right) barrier,
Fig.~1. The preexponential factor in Eq.~(\ref{tran}) takes into account a
mismatch between the effective masses, $m_{\sigma }$ and $m_{\ast }$, and
the velocities, $v_{\sigma x}$ and $v_{x}$, of electrons at the FM-S
interfaces (cf. Ref.\cite{Brat}). We consider $qV_{b}\lesssim \Delta _{0}$, $%
T<\Delta _{0}\ll \Delta $ and $E\geq E_{c}>F+T$, when\ Eqs.~(\ref{JRL}),(\ref
{tran}) yield the following result for the tunneling-emission current density

\begin{eqnarray}
j_{\sigma }^{L(R)} &=&\frac{\alpha qM_{c}T^{5/2}(8m_{\ast })^{1/2}v_{0\sigma
(\sigma ^{\prime })}^{L(R)}\exp (-\eta \kappa _{0}^{L(R)}l^{L(R)})}{\pi
^{3/2}\hbar ^{3}\left[ (v_{\sigma (\sigma ^{\prime
})}^{L(R)})^{2}+(v_{t0}^{L(R)})^{2}\right] }  \nonumber \\
&&\times \left( e^{\frac{^{F_{\sigma }^{L(R)}-E_{c}}}{T}}-e^{\frac{%
F_{L(R)}-E_{c}}{T}}\right) ,  \label{J}
\end{eqnarray}
where $\kappa _{0}^{L(R)}\equiv 1/l_{0}^{L(R)}=(2m_{\ast }/\hbar
^{2})^{1/2}(\Delta -\Delta _{0}\pm qV_{L(R)})^{1/2}$, $v_{t0}^{L(R)}=\sqrt{%
2(\Delta -\Delta _{0}\pm qV_{L(R)})/m_{\ast }},$ and $v_{\sigma (\sigma
^{\prime })}^{L(R)}=v_{\sigma (\sigma ^{\prime })}(\Delta _{0}\pm qV_{L(R)})$%
. It follows from Eqs.~(\ref{J}) and (\ref{nc}) that the spin currents of
electrons with the quantization axis $\parallel \vec{M}_{1}$ in FM$_{1}$
with $\sigma =\uparrow (\downarrow )$, Fig.~1$,$ and $\parallel \vec{M}_{2}$
in FM$_{2}$ with $\sigma ^{\prime }=\pm $ through the junctions of unit area
are equal to 
\begin{eqnarray}
J_{\sigma }^{L} &=&J_{0}^{L}d_{\sigma }^{L}\left( e^{\frac{qV_{L}}{T}%
}-2n_{\sigma }(0)/n\right) ,  \label{1} \\
J_{\sigma ^{\prime }}^{R} &=&J_{0}^{R}d_{\sigma ^{\prime }}^{R}\left(
2n_{\sigma ^{\prime }}(w)/n-e^{-\frac{qV_{R}}{T}}\right) ,  \label{2} \\
J_{0}^{L(R)} &=&-\alpha _{0}^{L(R)}nqv_{T}\exp (-\eta \kappa
_{0}^{L(R)}l^{L(R)}).  \label{J0}
\end{eqnarray}
Here we have introduced $\alpha _{0}^{L(R)}=1.6\left( \kappa
_{0}^{L(R)}l^{L(R)}\right) ^{1/3}$, the thermal velocity $v_{T}\equiv \sqrt{%
3T/m_{\ast }}$, and the spin factors $d_{\sigma }^{L}=v_{T}v_{\sigma }^{L}%
\left[ \left( v_{t0}^{L}\right) ^{2}+\left( v_{\sigma }^{L}\right) ^{2}%
\right] ^{-1}$ and $d_{\sigma ^{\prime }}^{R}=v_{T}v_{\sigma ^{\prime }}^{R}%
\left[ \left( v_{t0}^{R}\right) ^{2}+\left( v_{R\sigma ^{\prime }}\right)
^{2}\right] ^{-1}$.

Now we can find the dependence of current on a magnetic configuration in FM
electrodes and an external magnetic field. The spatial distribution of
spin-polarized electrons is determined by the kinetic equation $dJ_{\sigma
}/dx=q\delta n_{\sigma }/\tau _{s}$, where $\delta n_{\sigma }=n_{\sigma
}-n/2$, $\tau _{s}$ is spin-coherence lifetime of the electrons in the $n-$%
semiconductor, and the current in spin channel $\sigma $ is given by 
\begin{equation}
J_{\sigma }=q\mu n_{\sigma }E+qDdn_{\sigma }/dx,  \label{Js}
\end{equation}
where\ $D$ and $\mu $ are diffusion constant and mobility of the electrons, $%
E$ the electric field \cite{aronov76pik,Sze}. From conditions of continuity
of the total current, $J=J_{\uparrow }+J_{\downarrow }={\rm const}$ and $%
n=n_{\uparrow }+n_{\downarrow }={\rm const}$, it follows that $E(x)=J/q\mu n=%
{\rm const}$ and $\delta n_{\uparrow }=-\delta n_{\downarrow }$. Note that $%
J<0$, thus $E<0$. With the use of the kinetic equation and (\ref{Js}), we
obtain the equation for $\delta n_{\uparrow }(x)$ \cite{aronov76pik,flat}.
Its general solution is 
\begin{equation}
\delta n_{\uparrow }(x)=(n/2)(c_{1}e^{-x/L_{1}}+c_{2}e^{-(w-x)/L_{2}}),
\label{nx}
\end{equation}
where $L_{1(2)}=(1/2)\left( \sqrt{L_{E}^{2}+4L_{s}^{2}}+(-)L_{E}\right) $, $%
L_{s}=\sqrt{D\tau _{s}}$ and $L_{E}=\mu |E|\tau _{s}$ are the spin diffusion
and drift lengths \cite{aronov76pik,flat}. Substituting Eq.~(\ref{nx}) into (%
\ref{Js}), we obtain

\begin{equation}
J_{\uparrow }(x)=(J/2)\left[
1+b_{1}c_{1}e^{-x/L_{1}}+b_{2}c_{2}e^{-(w-x)/L_{2}}\right]  \label{J01}
\end{equation}
where $b_{1(2)}=L_{1}/L_{E}$ ($-L_{2}/L_{E})$. We consider the case when $%
w\ll L_{1}$ and the transit time $t_{tr}\simeq w^{2}/(D+\mu |E|w)$ of the
electrons through the $n-$semiconductor layer is shorter than $\tau _{s}$.
In this case spin ballistic transport takes place, i.e. the spin of the
electrons injected from the FM$_{1\text{\ }}$layer is conserved in the
semiconductor layer, $\sigma ^{\prime }=\sigma $. Probabilities of the
electron spin $\sigma =\uparrow $ to have the projections along $\pm \vec{M}%
_{2}$ are $\cos ^{2}\left( \theta /2\right) $ and $\sin ^{2}(\theta /2)$,
respectively, where $\theta $ is the angle between vectors $\sigma =\uparrow 
$ and $\vec{M}_{2}$. Therefore, the spin current through the right junction
can be written, using Eq.~(\ref{2}), as 
\begin{eqnarray}
J_{\uparrow (\downarrow )}^{R} &=&J_{0}^{R}\left[ 2n_{\uparrow (\downarrow
)}(w)/n-\exp (-qV_{R}/T)\right]  \nonumber \\
&&\times \left[ d_{+(-)}\cos ^{2}(\theta /2)+d_{-(+)}\sin ^{2}(\theta /2)%
\right] .  \label{JLR}
\end{eqnarray}

It follows from Eqs.~(\ref{1}) and (\ref{JLR}) that the total current $%
J=J_{\uparrow }^{L}+J_{\downarrow }^{L}=J_{\uparrow }^{R}+J_{\downarrow
}^{R} $ through the left and right interfaces is equal, respectively, 
\begin{eqnarray}
J &=&J_{0}^{L}(d_{\uparrow }+d_{\downarrow })[\gamma _{L}-2P_{L}\delta
n_{\uparrow }(0)/n],  \label{JL} \\
J &=&J_{0}^{R}(d_{-}+d_{+})[\gamma _{R}+2P_{R}\cos \theta \delta n_{\uparrow
}(w)/n],  \label{JR}
\end{eqnarray}
where $\gamma _{L}=e^{qV_{L}/T}-1$ and $\gamma _{R}=1-e^{-qV_{R}/T}$, and
\begin{eqnarray}
J_{\uparrow }^{L} &=&\frac{J}{2}\frac{(1+P_{L})\left[ \gamma _{L}-2\delta
n_{\uparrow }(0)/n\right] }{\gamma _{L}-2P_{L}\delta n_{\uparrow }(0)/n},
\label{J-d} \\
J_{\uparrow }^{R} &=&\frac{J}{2}\frac{(1+P_{R}\cos \theta )\left[ \gamma
_{R}+2\delta n_{\uparrow }(w)/n\right] }{\gamma _{R}+2P_{R}\cos \theta
\delta n_{\uparrow }(w)/n}.  \label{J-0}
\end{eqnarray}
Here we have\ introduced the spin polarization $P_{L(R)}=\left( d_{\uparrow
}^{L(R)}-d_{\downarrow }^{L(R)}\right) \left( d_{\uparrow
}^{L(R)}+d_{\downarrow }^{L(R)}\right) ^{-1}$ for the left (right) contact,
which is equal to 
\begin{equation}
P_{L(R)}=\frac{\left( v_{\uparrow }^{L(R)}-v_{\downarrow }^{L(R)}\right) %
\left[ (v_{t0}^{L(R)})^{2}-v_{\uparrow }^{L(R)}v_{\downarrow }^{L(R)}\right] 
}{\left( v_{\uparrow }^{L(R)}+v_{\downarrow }^{L(R)}\right) \left[
(v_{t0}^{L(R)})^{2}+v_{\uparrow }^{L(R)}v_{\downarrow }^{L(R)}\right] }.
\label{PRL}
\end{equation}
Importantly,{\bf \ }this $P_{L(R)}$ is determined by the electron states in
FM {\em above} the Fermi level, at $E=E_{c}>F$, which may be substantially
more polarized compared to the states at the Fermi level \cite{BO03}. In
general, the parameters $c_{1(2)},$ $V_{L(R)}$ are determined by
Eqs.~(\ref{J01}), (\ref{1}), (\ref{2}), and (\ref{JL})-(\ref{J-0}).\ 

The current through the structure is ohmic at small bias, $J\propto V$ at $%
\left| qV\right| <T,$  and saturates at larger bias
voltages. Indeed, from Eqs.~(\ref{J-d}),(\ref{J-0}) at $qV_{R(L)}\gtrsim 2T$
we find $J_{\uparrow }^{L}=\frac{J}{2}(1+P_{L})$ and 
\begin{equation}
J_{\uparrow }^{R}=\frac{J}{2}\frac{(1+P_{\theta })(1+2\delta n_{\uparrow
}(w)/n)}{1+2P_{\theta }\delta n_{\uparrow }(w)/n}.  \label{J22}
\end{equation}
where $P_{\theta }\equiv P_{R}\cos \theta $. We can obtain the current at $%
x=0$ ($w)$ from Eq.~(\ref{J01}), equate it to $%
J_{\uparrow }^{L}$ ($J_{\uparrow }^{R})$ assuming negligible spin
relaxation at the interface, and then find the unknown $c_{1(2)}$ with
the use of (\ref{nx}). 
At $L_{E}\gg L_{s}$, we have $%
b_{1}=1,$ $b_{2}=-L_{s}^{2}/L_{E}^{2}$ and $c_{1}=P_{L},\quad
c_{2}=-P_{\theta }(1-P_{L}^{2})/(1-P_{L}P_{\theta })$. Thus, according to
Eqs.~(\ref{nx}), the spin densities at the two interfaces are 
\begin{eqnarray}
2\delta n_{\uparrow }(0)/n &=&P_{L}-e^{-w/L_{2}}P_{\theta
}(1-P_{L}^{2})/(1-P_{L}P_{\theta }),  \label{n-0} \\
2\delta n_{\uparrow }(w)/n &=&\left( P_{L}-P_{\theta }\right) /\left(
1-P_{L}P_{\theta }\right).  \label{nw}
\end{eqnarray}

The spin-polarized density profile $n_{\sigma }(x)$ is shown in Fig.~2
(bottom panel)\ for $w\ll L_{s}$. One can realize from Eqs.~(\ref{nx}),(\ref
{n-0}) and (\ref{nw}) that large accumulation of the majority injected spin
occurs when the moments on the magnetic electrodes are antiparallel, $\vec{M}%
_{1}\parallel -\vec{M}_{2}$, and relatively small accumulation occurs in the
case of the parallel configuration, $\vec{M}_{1}\parallel \vec{M}_{2}$,
Fig.~2.

%%  FIGURE 2    %%%%%%%%%%%%%%%%%%%%%%%%%%%%%%%%%%%%%%%%%%%%%%%%%%%%%%%%%%%%
\begin{figure}[t]
\epsfxsize=3.2in \epsffile{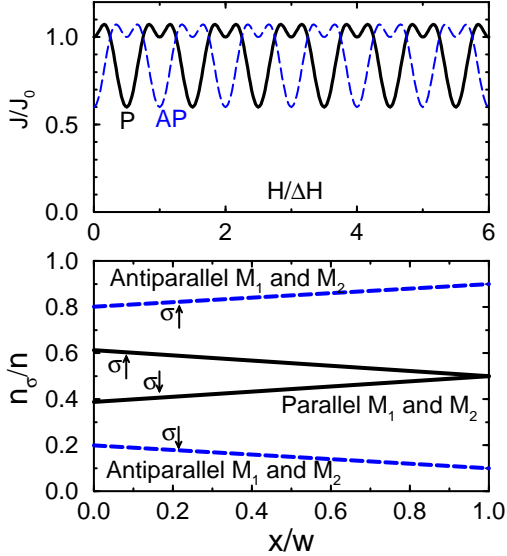}
\caption{(Color online) Oscillatory dependence of the current $J$ through the structure on
the magnetic field $H$ (top panel) for parallel (P)\ and antiparallel (AP)
moments $M_{1}$ and $M_{2}$ on the electrodes, Fig.~1, and $P_{L}=P_{R}=0.5.$
Spatial distribution of the spin polarized electrons $n_{\uparrow
(\downarrow )}/n$ in the structure for different configurations of the
magnetic moments $M_{1}$ and $M_{2}$ in the limit of saturated current
density $J$, $w=60$ nm, $L_{2}=100$ nm (bottom panel).}
\label{fig:fig2}
\end{figure}

At $qV>T$ the current saturates at the value 
\begin{equation}
J=J_{0}\left( 1-P_{R}^{2}\cos ^{2}\theta \right) \left( 1-P_{L}P_{R}\cos
\theta \right) ^{-1},  \label{Jtet}
\end{equation}
where $J_{0}=J_{0}^{R}(d_{+}+d_{-})$,
as follows from Eqs.~(\ref{n-0}) and (\ref{JR}). 
For the {\em opposite} bias, $qV<-T,$
the total current $J$ is given by Eq.~(\ref{Jtet}) with the replacement $%
P_{L}\rightleftharpoons P_{R}.$ The current $J$ is minimal for antiparallel (%
{\rm AP})$\;$moments $\vec{M}_{1}$ and $\vec{M}_{2}$ in the electrodes when $%
\theta =\pi $ and near maximal for parallel ({\rm P}) magnetic moments $\vec{%
M}_{1}$ and $\vec{M}_{2}$. The ratio $\frac{J_{\max }(P)}{J_{\min }(AP)}=%
\frac{1+P_{L}P_{R}}{1-P_{L}P_{R}}$ is the same as for the tunneling FM-I-FM
structure \cite{Slon,Brat}, hence, the structure may be also used as a
memory cell.

The present heterostructure has an additional degree of freedom, compared to
tunneling FM-I-FM structures, which can be used for a {\em magnetic} {\em %
sensing}. Indeed, spins of the injected electrons can precess in an external
magnetic field $H$ during the transit time $t_{tr}$ of the electrons through
the semiconductor layer ($t_{tr}<\tau _{s}$). In Eqs.~(\ref{JL}), (\ref{Jtet}%
) the angle between the electron spin and the magnetization $\vec{M}_{2}$ in
the FM$_{2}$ layer is in general $\theta =\theta _{0}+\theta _{H},$ where $%
\theta _{0}$ is the angle between the magnetizations $M_{1}$ and $M_{2},$
and $\theta _{H}$ is the spin rotation angle. The spin precesses with a
frequency $\Omega =\gamma H,$ where $H$ is the magnetic field normal to the
spin direction, and $\gamma =qg/(m_{\ast }c)$ the gyromagnetic ratio, $g$
the $g-$factor. Therefore, $\theta _{H}=\gamma _{0}gHt_{tr}(m_{0}/m_{\ast
}),$ where $m_{0}$ the mass of a free electron, $\gamma _{0}g=1.76\times
10^{7}$ Oe$^{-1}$c$^{-1}$ for $g=2$ (in some magnetic semiconductors $g\gg 1$%
). According to Eq.~(\ref{Jtet}), with increasing $H$\ the current {\em %
oscillates} with an amplitude $(1+P_{L}P_{R})/(1-P_{L}P_{R})$ and period $%
\Delta H=(2\pi m_{\ast })(\gamma _{0}gm_{0}t_{tr})^{-1}$, Fig.~2 (top
panel). Study of the current oscillations at various bias voltages allows to
find $P_{L}$ and $P_{R}$.

For magnetic sensing one may choose $\theta _{0}=\pi /2$ ($\vec{M}_{1}\perp 
\vec{M}_{2}$). Then, it follows from Eq.~(\ref{Jtet}) that for $\theta
_{H}\ll 1$%
\begin{eqnarray}
J &=&J_{0}[1+P_{L}P_{R}\gamma _{0}gHt_{tr}(m_{0}/m_{\ast })]=J_{0}+J_{H},
\label{H} \\
K_{H} &=&dJ/dH=J_{0}P_{L}P_{R}\gamma _{0}gt_{tr}(m_{0}/m_{\ast }),
\label{Gain}
\end{eqnarray}
where $K_{H}$ is the magneto-sensitivity coefficient. For example, $%
K_{H}\simeq 2\times 10^{-3}J_{0}P_{L}P_{R}$ A/Oe for $m_{0}/m_{\ast }=14$
(GaAs) and $g=2$, $t_{tr}\sim 10^{-11}$s, and the angle $\theta _{H}=\pi $
at $H\simeq 1$ kOe. Thus, $J_{H}\simeq 1$ mA at $J_{0}=25$ mA, $%
P_{L}P_{R}\simeq 0.2,$ and $H\simeq 100$ Oe. The maximum operating speed of
the field sensor is very high, since redistribution of nonequilibrium
injected electrons in the semiconductor layer occurs over the transit time $%
t_{tr}=w/\mu |E|=J_{s}w\tau _{s}/\left( JL_{s}\right) $, $t_{tr}\lesssim
10^{-11}$s for $w\lesssim 200$ nm, $\tau _{s}\sim 3\times 10^{-10}$s, and $%
J/J_{s}\gtrsim 10$ ($D\approx 25$ cm$^{2}$/s at $T\simeq 300$ K \cite{Sze}).
Thus, the operating frequency $f=1/t_{tr}\gtrsim 100$ GHz ($\omega =2\pi
/t_{tr}\simeq 1$ THz) would be achievable at room temperature.

We emphasize that the parameters $\kappa _{0}^{L(R)}$, $P_{L(R)}$ are
functions of the bias $V_{L(R)}$ and $\Delta _{0}$. The efficient spin
injection can be achieved when the bottom of conduction band in a
semiconductor $E_{c}$ near both FM-S junctions is close to a peak in a
density of spin polarized states, e.g. of minority electrons in the
elemental ferromagnet like Fe, Co, Ni (cf. \cite{BO03}). For instance, in Ni
and Fe the peak is at $F+\Delta _{\downarrow }$, $\Delta _{\downarrow
}\simeq 0.1$ eV \cite{Mor}.

In conclusion, we have showed that (i) the present heterostructure can be
used as a sensor for an ultrafast nanoscale reading of an inhomogeneous
magnetic field profile, (ii) it includes two FM-S junctions and can be used
for measuring spin polarizations of these junctions, and (iii) it is a {\em %
multifunctional} device where current depends on mutual orientation of the
magnetizations in the ferromagnetic layers, an external magnetic field, and
a (small) bias voltage, thus it can be used as a logic element, a magnetic
memory cell, or an ultrafast read head.

%%%%% References: %%%%%

\end{document}